\documentclass[twocolumn,showpacs,reprint,amsmath,amssymb,aps,prapplied,longbibliography,superscriptaddress]{revtex4-2}
\usepackage{color}
\usepackage{graphicx}
\usepackage{textcomp} 
\usepackage{subeqnarray}
\usepackage[caption=false]{subfig}

\begin{document}

\title{Valveless pumping at low Reynolds numbers}

\author{Gabriel Amselem}\email[Corresponding author: ]{amselem@ladhyx.polytechnique.fr}
\author{Christophe Clanet}
\author{Michael Benzaquen}
\affiliation{Laboratoire d'Hydrodynamique, CNRS, \'{E}cole polytechnique, Institut Polytechnique de Paris, 91120 Palaiseau, France}

\date{\today}
\begin{abstract}
 Pumping at low Reynolds number is a ubiquitously encountered feature, both in biological organisms and   engineered devices. Generating net flow requires the presence of an asymmetry in the system, which traditionally comes from geometric flow rectifiers. Here, we study a valveless system of $N$ oscillating pumps in series, where the asymmetry comes not from the geometry but from time, that is the phase shifts between the pumps. Experimental and theoretical results are in very good agreement. We provide the optimal phase shifts leading to the maximal net flow in the continuous $N\rightarrow \infty$ limit. The maximal flow rate is larger by 25\% than that of a traditional peristaltic sinusoidal wave, for the same amplitude of actuation. Our results pave the way for the design of more efficient microfluidic pumps. 
\end{abstract}
\maketitle

\section{Introduction}

Pumping at low Reynolds number is a broadly encountered feature,  both in nature and engineered systems. On the biological side, insects such as butterflies and mosquitoes feed essentially on liquids such as plant nectar or blood, and many rely on the action of one or more muscular pumps to intake their food~\cite{borrell2006}. On the man-made side, the design of micropumps for MEMS applications started with the seminal work of Smits 30 years ago~\cite{smits1990}, and continued with the Quake-valves now ubiquitous in microfluidic designs~\cite{unger2000}, triggering a large variety of novel designs~\cite{laser2004,woias2005,iverson2008,bussmann2021}.

In one way or another, these natural and engineered devices rely on the presence of an asymmetry in the system, which is essential to pump at low Reynolds number.  Some biological and man-made pump designs rely on valves, a role played by the pharynx in mosquitoes~\cite{uchida1979,kim2012,kikuchi2018}. Yet, valves are prone to mechanical failure and clogging, and valveless systems of micropumps were conceived to make them easier to fabricate and more resilient~\cite{woias2005,stemme1993}. Their mode of operation relies on geometrically asymmetric elements having flow-rectifying properties. Valveless pumping at low Reynolds number is also prevalent in biology, throughout the animal kingdom: in the early stages of embryonic development of a number of species, the heart consists in two chambers that pump fluid without a valve~\cite{glickman2002cardiac,olson2006gene,scherz2008high}.

In this communication we study experimentally and theoretically a valveless pumping system consisting of $N$ identical contracting  pumps in series, operating at low Reynolds number. The system does not contain any flow rectifier  (valveless), and the asymmetry comes not from the built-in geometry but from the phase shifts between the oscillating pumps, i.e. from the dynamics of actuation. The average flow rate through the system is computed from the system geometry and  the phases of the oscillating pumps. Theoretical results are in very good agreement with experiments. We find numerically the phase shifts between the $N$ pumps leading  to maximal flow rate, for arbitrary values of $N$. When the number of pumps is increased to infinity (continuous limit), and for a given amplitude of deformation, the maximum flow rate obtained is larger by 25\% than that of a traditional peristaltic, sinusoidal wave. 
Our results thereby cut the first turf for more efficient designs of microfluidic pumps.

\begin{figure}[h!]
 \centering
 \includegraphics[width=0.8\columnwidth]{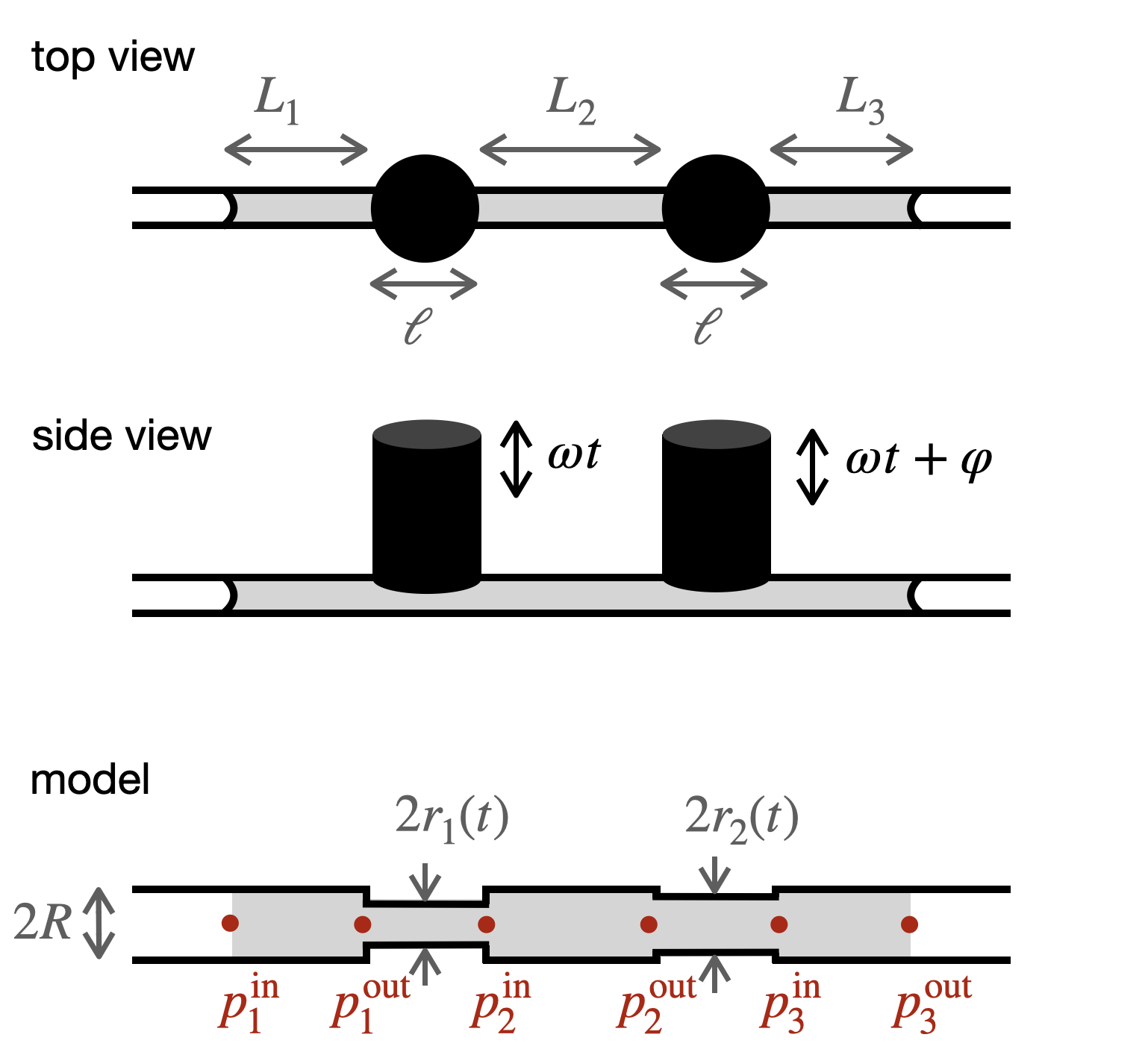}
 \caption{Sketch of the experimental setup: two cylindrical wooden rods of diameter $1\; \rm cm$ placed on computer-controlled syringe pumps are used to press periodically on an elastic silicone tubing. The tubing is partially filled with V1000 silicon oil (gray). The parameterization of the system is as follows: two pumps of length $\ell$ delimit sections of a tubing of total length $L_{\rm tot} = 2\ell+ L_1+L_2+L_3$. The tubing radius is $R=0.75\; \rm mm$. The varying tubing radius under the pumps is denoted $r_{1,2}(t)$. }
 \label{fig:setup}
\end{figure}

\begin{figure*}[ht!]
 \centering
 \includegraphics{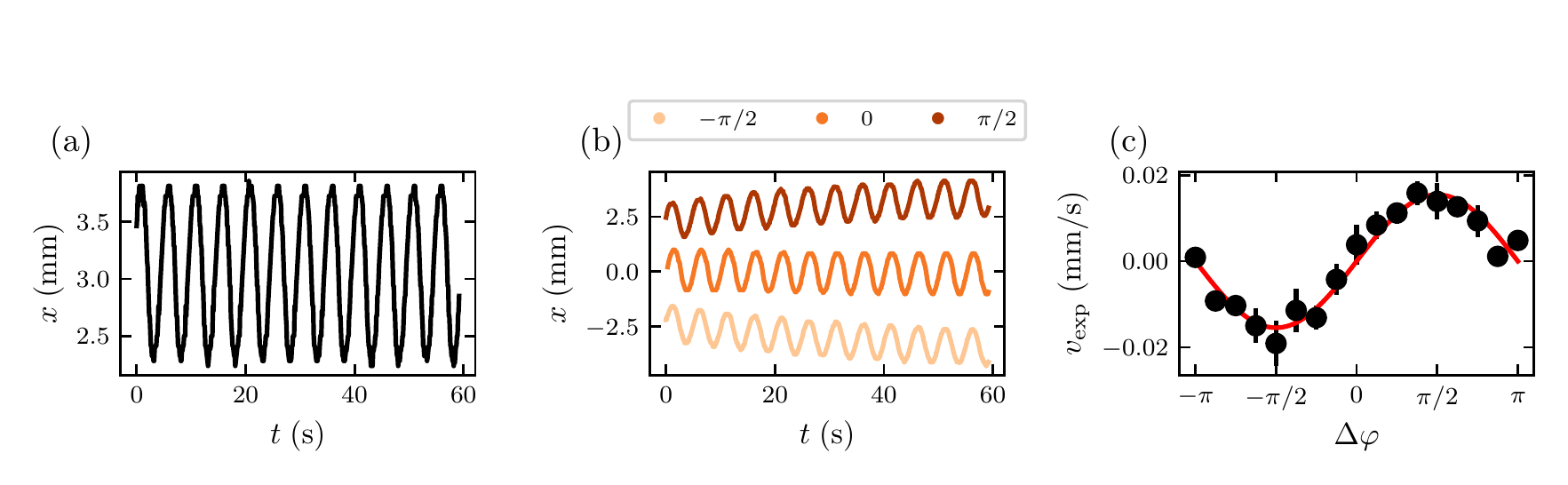}
 \caption{(a) Oscillating position of the oil/air meniscus when one pump presses periodically on the silicone tubing, with a period $T=5\; \rm s$.  (b) When two pumps press periodically on the silicone tubing, the position of the meniscus shows a net drift superposed on oscillations at the forcing period. The drift direction depends on the sign of the phase shift $\Delta\varphi = \varphi_1 - \varphi_2$ between the two pumps. (c) Evolution of the mean drift speed $v_{\rm exp}$ as a function of the phase shift $\Delta\varphi$. Points: experimental data. Line: fit to a sine function.}
 \label{fig:traj}
\end{figure*}

\section{Pumping with two pumps}
\subsection{Experimental setup} 

The experimental setup consists in a silicon tubing of internal diameter $2R = 1.5$\,mm (Thermo Scientific TM Sterilin, Thermo Fischer, France), partially filled with silicon oil (Rhodorsil\,\textsuperscript{\tiny\textregistered} 47V1000, kinematic viscosity $\nu = 10^{-3} \; \rm{m^2\cdot s^{-1}}$). Cylindrical wooden rods of diameter 1\,cm are placed on computer-controlled syringe pumps (Nemesys, Cetoni GMBH, Germany), and press periodically on the silicon tubing, see Fig.~\ref{fig:setup}. These wooden rods constitute the pumps. The evolution of the tubing diameter at the location of rod $i$ can be approximated as $r_i(t) = R[1-\frac{\varepsilon}{2} + \frac{\varepsilon}{2} \sin(\omega t + \varphi_i)]$, where $\varepsilon$ is the normalized pumping amplitude, $\omega$ the pumping angular frequency and $\varphi_i$ the phase shift  of pump $i$, see Fig.~\ref{fig:setup}. The column of oil of length $L_{\rm tot}$ is imaged at 10~Hz with a CMOS camera (Lt225c, Teledyne Lumenera, Canada), and the position of the oil/air interface is followed over time. 

When only one pump is actuated, the oil/air meniscus moves periodically and symmetrically left and right at the forcing frequency, leading to on average zero net flow through the system, see Fig.~\ref{fig:traj}a. This is typical of the reversible behavior of flows at low-Reynolds number: in the absence of inertia, a symmmetric forcing on the fluid does not lead to any net flow~\cite{purcell1977}. To estimate the value of $\varepsilon$, we measure the length of the column of liquid $L_{\rm tot, 0}$ in the absence forcing. We then measure its length $L_{\rm tot, max}$ when a single pump is actuated, and maximally pressing on the tube. The resulting effective radius of the deformed cross-section is then $r_{\rm min} = r(1-\varepsilon)$. By volume conservation, we find that $\varepsilon = 1 - \sqrt{1-(L_{\rm tot,max} - L_{\rm tot, 0})/\ell}$, where $\ell=1\; \rm cm$ is the diameter of the wooden rod, and so the size of the pumping region.

When two pumps are actuated, the meniscus still moves periodically left and right at the forcing frequency, but in the presence of an additional, slower drift. The amplitude and direction of the drift depend on the phase shift $\Delta\varphi = \varphi_1 - \varphi_2$ between the two pumps, see Fig.~\ref{fig:traj}b. The drift velocities of the meniscus are extracted from kymographs of the experiment, for all values of the phase shifts tested. The drift velocity $v$ shows a sinusoidal dependency on the phase shift $\Delta\varphi$, see Fig.~\ref{fig:traj}c, with no drift for $\Delta \varphi=0, \pm \pi$, and a maximum drift for $\Delta\varphi=\pm \pi/2$. This can be intuitively understood by seeing that when the two pumps are in phase opposition ($\Delta\varphi=\pm \pi$), the fluid pushed by one pump fills up the space of the tube under the second pump, resulting in a reversible situation and so an absence of net flow. When the phase shift is  $\Delta\varphi=\pm \pi/2$, the pump in phase retardation pushes the fluid preferentially away from the other pump, in a discretized version of peristaltic pumping.

\subsection{Theory} 
The flow generated by the pumps can be calculated under the assumption of small deformations $\varepsilon \ll 1$.  Consider the valveless system of two pumps pushing on a tubing of radius $R$ and total length $L_{\rm tot}$, as shown in Fig.~\ref{fig:setup}. The pumps have a length $\ell$, and the radius of the tubing at the location of the pumps is $r_i(t)$, $i=1,2$. The tubing is delimited by the pumps into three sections of length $L_1$, $L_2$ and $L_3$. The total length of the tubing is then $L_{\rm tot} = L_1 + L_2+L_3 + 2\ell$. The pressures at the inlet and outlet of the $i$-th section are $p_i^{\rm in}(t)$ and $p_i^{\rm out}(t)$, respectively (see bottom panel in Fig.~\ref{fig:setup}). At low Reynolds numbers and in the cylindrical geometry presented here, flows obey Poiseuille's law, so that the flow rate $Q_i(t)$ in channel $i$ is linked to the pressures $p_i^{\rm in}(t)$ and $p_i^{\rm out}(t)$ through:
\begin{equation}
\label{eqQ2pumps}
\forall i \in [1, 3], \qquad Q_i(t) = k_i[p_i^{\rm in}(t)-p_i^{\rm out}(t)], 
\end{equation}
where $k_i = \pi R^4/8\eta L_i$ is the resistivity of the channel. The flow rates between adjacent channels are related by mass conservation:
\begin{equation}
\label{eqDiffQ2pumps}
\forall i \in [1, 2] , \qquad Q_i(t) - Q_{i+1}(t) = 2 \ell \pi r_i(t)  \partial_t r_i.
\end{equation}
Last, the flow in channel $i$ can be expressed as a function of the resistivity $\kappa_i(t) = \pi r_i(t)^4/8\eta \ell$ of pump $i  \in [1, 2]$:
\begin{equation}
\label{eqQc2pumps}
  Q_i(t)  = {\kappa_i}(t)[p_i^{\rm out}(t)-p_{i+1}^{\rm in}(t)] + \ell \pi r_i(t) \partial_t r_i.
\end{equation}
Equations~\eqref{eqQ2pumps}-\eqref{eqQc2pumps} constitute a linear system of 7 scalar equations. The pressures are imposed to be identical at the inlet and outlet of the system: $p_1^{\rm in} = p_3^{\rm out} = p_0$. We solve for the $7$ unknown variables $(Q_1,Q_2, Q_3, p_1^{\rm out}, p_2^{\rm in}, p_2^{\rm out} ,p_3^{\rm in})$,  looking for a solution to the equations when pumps are forced sinusoidically: $r_i(t) = R \left [1 -\frac{\varepsilon}{2} + \frac{\varepsilon}{2} \sin(\omega t + \varphi_i) \right ]$.

The time-averaged flow $\bar Q_1 = \bar Q_2 =  \bar Q_3  :=   \bar Q$ is found to depend on the phase difference $\Delta\varphi = \varphi_1 - \varphi_2 $ between the two pumps through: 
\begin{equation}
\label{eq:avgFlux2Pumps}
\bar Q(\Delta\varphi) = \frac{\pi R^2 \ell^2(L_{\rm tot}-\ell-L_2)}{L^2_{\rm tot}} \omega \varepsilon^2 \sin \Delta\varphi + \mathcal{O}(\varepsilon^3).
\end{equation}
 
The average flow velocity $v_{\rm theor, \; 2 \; pumps} = \bar{Q}/\pi R^2$ in the tubing is then given by:
\begin{equation}
\label{eq:avgSpeed2Pumps}
v_{\rm theor, \; 2 \; pumps} = (L_{\rm tot}-\ell-L_2)\frac{\ell^2}{L^2_{\rm tot}} \omega \varepsilon^2 \sin \Delta\varphi + \mathcal{O}(\varepsilon^3).
\end{equation}

We recover the experimentally observed dependency of the flow speed on the sine of the phase difference $\sin\Delta\varphi$. The flow speed is proportional to the forcing frequency $\omega$, as can be expected for low-Reynolds  flows, and depends on a geometric prefactor that involves the ratio of the pump size $\ell$ to the total length of the fluid column $L_{\rm tot}$: the larger the fraction of the fluid being pressed by the pumps, the more efficient the pumping. Note that $\ell=L_{\rm tot}/2$ at most, for which the flow speed saturates to its maximum value. Last, note that pumping is a second-order effect in $\varepsilon$. 

\subsection{Comparison between experiments and theory}

\begin{figure}[ht!]
 \centering
 \includegraphics[width=\linewidth]{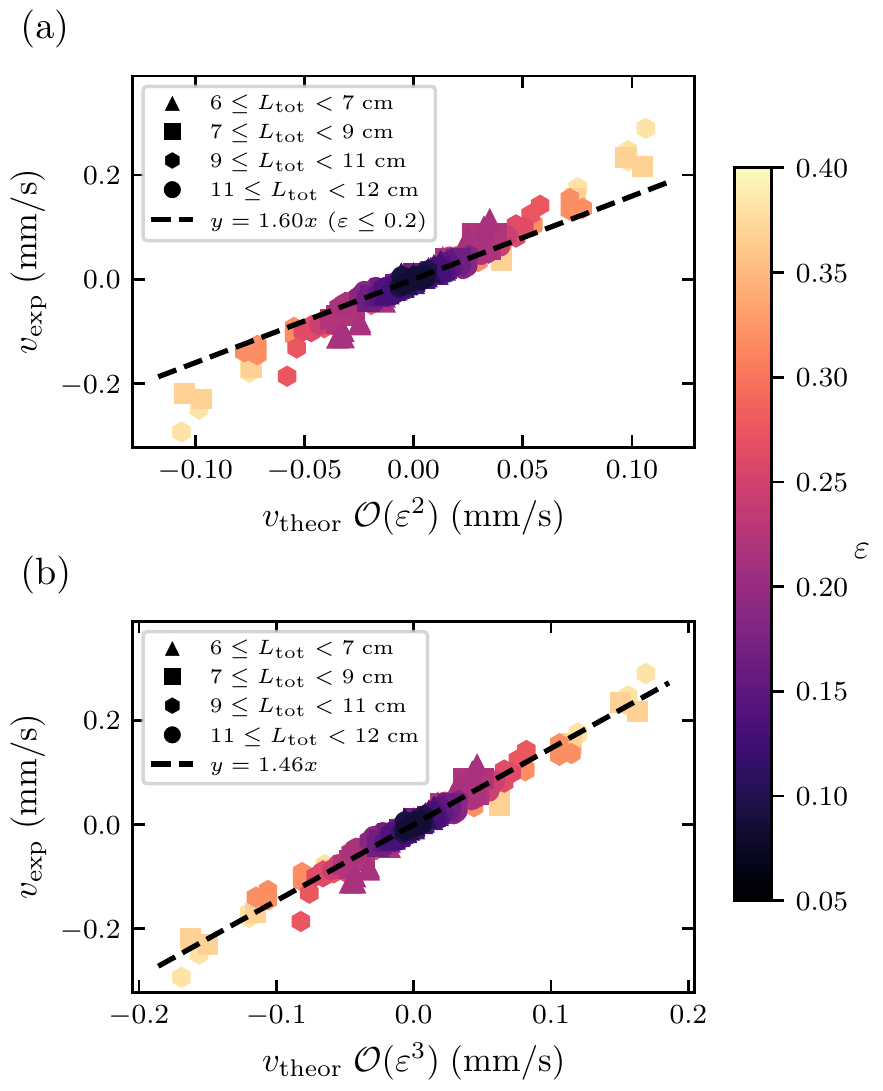}
 \caption{Comparison between the experimental and  theoretical flow speeds (points), for a system of two pumps. The normalized pumping amplitude $\varepsilon$ is color-coded, different symbols are for different lengths of liquid column $L_{\rm tot}$. Other parameters: $2R=1.5 \; \rm mm$, $l=1\; \rm cm$, $L_2=3\; \rm cm$. (a) Experimental vs. theoretical speed calculated up to $\mathcal{O}(\varepsilon^2)$. The dashed line is the best fit to data with $\varepsilon \leq 0.2$. (b) Experimental vs. theoretical speed calculated up to $\mathcal{O}(\varepsilon^3)$. The dashed line is the best fit to all data, and shows the line $y=1.46x$. For a discussion on the existence of a numerical prefactor in the linear fit of the data, see main text.}
 \label{fig:2pumps}
\end{figure}

Experiments are repeated for different values of $\varepsilon$, $L_{\rm tot}$ and $\Delta\varphi$, and the experimental drift velocities of the oil are plotted as a function of the theoretical velocities in Fig.~\ref{fig:2pumps}a.  We find good agreement in the scaling between the second-order theory and experiments for small values of $\varepsilon\lesssim 0.2$, see dashed line in Fig.~\ref{fig:2pumps}a. Deviation of the data from the dashed line at larger values of $\varepsilon$ indicate that higher-order terms are needed to fully describe the experimental data. Solving the equations of fluid motion up to order $\mathcal{O}(\varepsilon^3)$ and comparing the  experiments to the third-order theory gives very good agreement for all parameters tested, up to a numerical prefactor of $\approx 1.5$, see Fig.~\ref{fig:2pumps}b. The expression for the theoretical speed up to third order in $\varepsilon$ is given in \footnote{see Supplemental Material at [URL will be inserted by
 publisher], which includes Ref.~\cite{ramanujan1914modular}, for detailed expressions of the pumping speed, estimates of the tubing deformation and pumping energy, and a movie of the optimal waveform}\setcounter{footnote}{1}. 

The fact that the experimental speed is $\approx 1.5$ times larger than predicted by theory can most likely be attributed to the complex deformation of the tube when the pumps are actuated. Indeed, under an actuator, the cross-section of the tubing does not stay circular but becomes approximately elliptical, which leads to an increased hydrodynamic resistance at the location of the actuator~\cite{mortensen2005reexamination}. Such an increased resistance itself leads to a larger pumping efficiency. Moreover, due to elasticity, the deformation of the tubing extends beyond the region of the actuator, which also affects the hydrodynamic resistance of the tube. A discussion of these two effects can be found in~\cite{Note1}. Last, note that the resistance to flow is inversely proportional to the square of the cross-sectional area~\cite{mortensen2005reexamination}, and so a small error in the estimation of $\varepsilon$  can lead to large differences in the hydrodynamic resistance.

\begin{figure}[ht!]
 \centering
 \includegraphics[width=\linewidth]{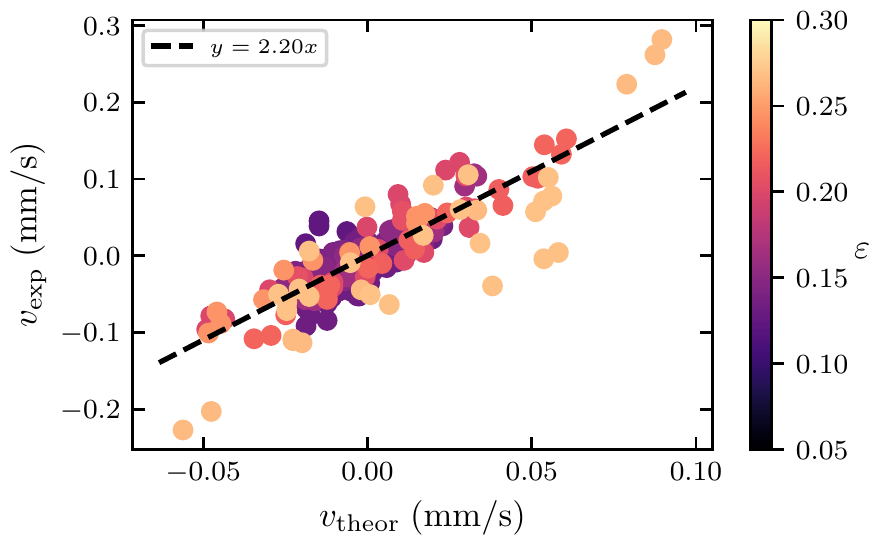}
 \caption{Comparison between the more than 200 experimental flow speed values measured and the theoretical speed values (points), for a system of three pumps. The normalized pumping amplitude $\varepsilon$ is color-coded. The liquid length $L_{\rm tot}$ was varied between 13 and 18 cm. Other parameters: $2R=1.5 \; \rm mm$, $l=1\; \rm cm$. Theoretical speed calculated up to $\mathcal{O}(\varepsilon^2)$, according to Eq.~\eqref{eq:avgSpeed3Pumps}. The dashed line is the best fit to all data. The existence of a numerical prefactor between theory and experiment is discussed in the main text.}
 \label{fig:3pumps}
\end{figure}

\section{Pumping with $N\geq 3$ pumps} 

\subsection{3 pumps}
The theory can be extended to a system of $N$ pumps in series (see below). In particular, for a system of 3 identical pumps, and using the same notations as above, the average flow velocity is found to be:
\begin{eqnarray}
  v_{\rm theor, \; 3\; pumps} &=& \frac{\ell^2}{L^2_{\rm tot}} \omega \varepsilon^2 \big[a_1\sin (\varphi_1 - \varphi_2) + a_2\sin (\varphi_1 - \varphi_3) \nonumber \\&& \qquad\qquad +\ a_3\sin (\varphi_2 - \varphi_3)\big] + \mathcal{O}(\varepsilon^3),
  \label{eq:avgSpeed3Pumps}
\end{eqnarray}
where the $a_i$ are geometrical prefactors that have the dimension of a length, and $\varphi_i$ is the phase shift of pump $i$ (see~\cite{Note1} for the expressions of the $a_i$). The validity of this prediction is tested by mounting a system with 3 pumps, where the forcing amplitude $\varepsilon$, the length of the column of oil $L_{\rm tot}$, and the phase shifts $\varphi_1$ and $\varphi_2$ are systematically varied. Again, experimental results are in good agreement with theoretical predictions, see Fig.~\ref{fig:3pumps}. As for the system with two pumps, the experimental velocities for the three-pump device are larger than theoretical predictions, by a factor $\approx 2$ (see black dashed line in Fig.~\ref{fig:3pumps}), likely for the same reasons as cited above.

\subsection{$N$ pumps} 
Is pumping more efficient when more pumps are added to the system? To answer this question,  consider a system of $N$ identical pumps of length $\ell$ in series, actuated sinusoidally. For simplicity, we now consider that all subsections of the tubing have the same length $L$, and so the same resistivity $k = \pi R^4/8\eta L$. The total length of the tubing is then $L_{\rm tot} = N\ell + (N+1) L$.  Similarly to Eq.~\eqref{eqQ2pumps}-\eqref{eqQc2pumps}, there are now $3N+1$ equations to solve in order to compute the flow rate: 
\begin{subeqnarray}
\slabel{eqQ}
&&\forall i \in [1, N+1], \quad Q_i(t) = k[p_i^{\rm in}(t)-p_i^{\rm out}(t)]. 
\\
\slabel{eqDiffQ}
 &&\forall i \in [1, N] , \quad Q_i(t) - Q_{i+1}(t) =  2 \ell \pi r_i(t)  \partial_t r_i,
\\
\slabel{eqQ2}
 && \quad \quad\quad Q_i(t)  = {\kappa_i}(t)[p_i^{\rm out}(t)-p_{i+1}^{\rm in}(t)] +  \ell \pi r_i(t)  \partial_t r_i.\quad\quad
\end{subeqnarray}
The same pressure is imposed at the inlet and the outlet ($p_1^{\rm in} = p_{N+1}^{\rm out} = p_0$), and the system of equations is solved for the $3N+1$ unknown variables $(Q_1,\cdots, Q_N, p_1^{\rm out}, p_2^{\rm in}, p_2^{\rm out},\cdots ,p_{N}^{\rm out}, p_{N+1}^{\rm in})$, when pumps are forced sinusoidically $r_i(t) = R \left [1 - \frac{\varepsilon}{2} + \frac{\varepsilon}{2}\sin(\omega t + \varphi_i) \right ]$.

The average flow rate over one period $T$ through a system of $N$ pumps is found to be: 
\begin{eqnarray}
\label{eq:QNPumps}
\bar Q(\{\varphi_i\}) &=& 4\pi R^2 \ell \omega  f_N(\varepsilon,b,\{\varphi_i\}) + \mathcal{O}(\varepsilon^3),
\end{eqnarray}
where we have introduced $b=\ell/L$, the ratio of the pump dimension to that of the tubing sections, and  
\begin{multline}
\label{eq:fn}
f_N (\varepsilon,b,\{\varphi_i\})=  \frac{ b}{(Nb+N+1)^2} \times \\ \left(\sum_{M=1}^{N-1}(M(b+1)+1)\sum_{i=1}^{M} \sin(\varphi_i-\varphi_{i+N-M})\right)\varepsilon^2 + O(\varepsilon^3) \ .
\end{multline}

The flow rate therefore depends on the phase shifts between the pumps, and on the aspect ratio $b=\ell/L$; it increases linearly with $b$ for small $b$ and saturates for $b \gg 1$, when the pumps essentially press on the entire length of the tubing. 

The phase shifts giving the maximal flow rate for a given aspect ratio $b$ were obtained by solving Eq.~\eqref{eq:fn} numerically, using Mathematica (Wolfram Research, Inc., IL, USA). For the particular case of $N=2$ pumps, the maximum flow rate is obtained for $\Delta \varphi=\pm \pi/2$, independent of the aspect ratio $b$.  
For a system of $N\geq 3$ pumps, the optimal phase shifts depend on the aspect ratio $b$, albeit only weakly (for $N=3$ pumps,  varying $b$ over 10 orders of magnitude changes the optimal values of the phase shifts by $\approx 5\%$, see Supp. Fig.~2 in~\cite{Note1}). Actually, for all tested values of $N\geq 3$, the optimal phases shifts are essentially constant for $b\ll 1$ and $b\gg 1$, and undergo a small variation when the aspect ratio $b$ goes from $0.1$ to $10$.

We now fix the aspect ratio $b=1$ as well as the tube size $L_{\rm tot}$, and consider that pumps of size $\ell = L_{\rm tot}/(2N+1)$ are actuated. The flow rate in the system is then simply given by $\tilde{Q}(\{\varphi_i\}) = \bar Q(\{\varphi_i\})/(2N+1)$. How does the optimal flow rate depend on the number of pumps $N$? We find that the maximal flow rate in the system $\tilde{Q}_{\rm max}(N)$ increases with the number of pumps $N$ until saturation in the continuous limit ($N\rightarrow\infty$), see black markers in Fig.~\ref{fig:nPumpsTheory}a. The maximal flow rate in a system of 2 pumps is $\approx 50\%$ as large as in the continuous limit. The increase in flow rate is very steep when $N$ initially increases: the flow rate in a system of 4 pumps is already $\approx 80\%$ that of the continuous limit. 

Regardless of the number of pumps $N$ in the system, the optimal flow rate $\tilde{Q}_{\rm max}$ is always larger than the flow rate $\tilde{Q}_{\rm sine}$ obtained by naively discretizing a sine wave, for which the phase shifts are equally spaced by $2\pi/N$ (see the orange markers in  Fig.~\ref{fig:nPumpsTheory}a, and the dashed orange line in Fig.~\ref{fig:nPumpsTheory}b). This is most obvious for $N=2$, where  discretizing a sine wave leads to the two pumps being actuated in phase opposition, which does not induce any net flow. For $N=3$ pumps, the optimal flow rate is $\approx 65\%$ larger than $\tilde{Q}_{\rm sine}$ (see inset in Fig.~\ref{fig:nPumpsTheory}a). The ratio $\tilde{Q}_{\rm max}/\tilde{Q}_{\rm sine}$ decreases when $N$ increases, and in the continuous limit, the optimal flow rate is 25\% larger than the flow rate obtained with a sinusoidal peristaltic wave, see the difference between black and orange points in Fig.~\ref{fig:nPumpsTheory}a and inset in Fig.~\ref{fig:nPumpsTheory}a. 

The waveform leading to the maximal flow rate in the continuous limit is shown in black in Fig.~\ref{fig:nPumpsTheory}b for large $N$, at time $t=0$. This optimal wave travels along the tube over one period of oscillation. The optimal waveform is different from the traditional sinusoidal peristaltic wave; in particular  the phase values of the first and last pumps are different, see the animated phase profiles in the Supplementary Movie~\cite{Note1}.

\section{Discussion} 

The literature on miniaturized pumps, which operate at low Reynolds number, is extremely vast, and covers aspects from  fluid dispensing in lab-on-a-chip devices for biological analysis~\cite{wang2018} to biomedical applications such as microdosing for drug delivery~\cite{bussmann2021}.  Yet, the broad majority of the existing work is concerned with the flow generated by at least 3 pumps in series, with very few exceptions on pumping with two pumps~\cite{berg2003a,geipel2007,kim2014}. A physical model of the flow due to pumping with 3 pumps was described in Goulpeau et al.~\cite{goulpeau2005}. In our work, we studied in detail the influence of the geometric parameters of the fluidic circuit, and of the phase shifts between the pumps, on the flow rate, for $N=2$ and 3 pumps, and extended the theory of pumping to an arbitrary number $N$ of pumps.

\begin{figure}[t!]
 \centering
 \includegraphics[width=0.98\columnwidth]{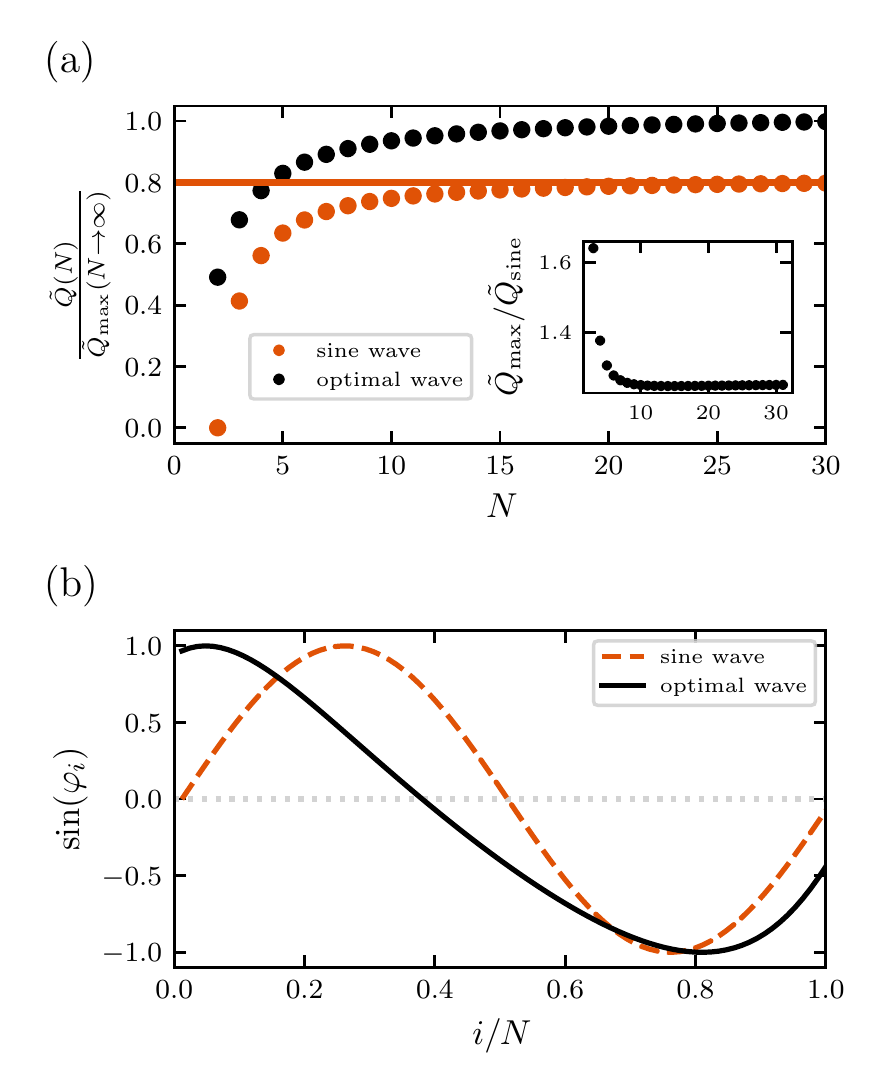}
 \caption{(a) Increasing the number of pumps in the system while keeping the tube length constant increases the optimal flow rate $\tilde{Q}_{\rm max}(N)$ (black dots), up to a limit corresponding to the continuous case $N\rightarrow\infty$. The optimal flow rate is always larger than the flow rate obtained by naively discretizing a sine wave, where the phase shifts are equally spaced by $2\pi/N$ (orange dots). Inset: ratio between the optimal flow rate $\tilde{Q}_{\rm max}(N)$, and the flow rate  $\tilde{Q}_{\rm sine}(N)$ obtained by discretizing a sine wave, as a function of $N$. (b) Waveform leading to the maximum flow in a system of $N$ pumps with $N=30$. Increasing $N$ further does not change the waveform.}
 \label{fig:nPumpsTheory}
\end{figure}

To achieve the highest flow rate possible, existing work usually sets the phase shifts to arbitrary, intuitive values, and then measures the flow rate as a function of the frequency of actuation of the pumps, see e.g.~\cite{xie2004,jang2007,liu2010,fuchs2012,kim2014}. At low Reynolds number however, both quantities are proportional; deviation from proportionality indicates either a change of regime to high Reynolds number, or reveals the presence of a compliance in the microsystem, which induces a phase shift between the applied actuation signal and the real mechanical actuation. This difference between the applied actuation and the real actuation is probably at the origin of surprising results, such as obtaining a maximum flow rate in a system of two pumps when they are supposedly actuated in phase ($\Delta\varphi = 0$) or in phase opposition ($\Delta\varphi = \pi$)~\cite{kim2014}, in direct contradiction with the reversibility of flows at low Reynolds number. Our results show that, in the low Reynolds regime, valveless pumping with two pumps is optimal for phase shifts between the pumps $\Delta\varphi = \pm\pi/2$, and we provide the exact analytical expression of pumping efficiency as a function of the geometric parameters of the problem. Theoretical and experimental results are in good agreement, see Figs.~\ref{fig:2pumps} and~\ref{fig:3pumps}.

The flow due to a simple peristaltic wave travelling along a
tube can be analytically calculated, see~\cite{shapiro1969,jaffrin1971}, with a number of applications. Original ideas making use of peristaltic waves were put forth, e.g. for the swimming of microorganisms~\cite{ajdari1994,ajdari1999}. The optimization of peristaltic wave shapes at low Reynolds number has been studied numerically using variational methods~\cite{walker2010} and, more recently, boundary integral methods~\cite{bonnet2020}. Here, we obtained numerical results on the optimal peristaltic wave shape at low Reynolds number by studying a system of $N$ discrete pumps, and let  $N\rightarrow\infty$, see Fig.~\ref{fig:nPumpsTheory}. The non-trivial optimal peristaltic shape we find leads to a flow rate larger by 25\% than that due to a sinusoidal peristaltic wave. Moreover, in the discrete case of a small number of pumps, the difference between the optimal flow rate and the flow rate obtained by naively discretizing a sine wave is even larger: for $N=3$ pumps, the optimal flow rate is $\approx 65\%$ larger than that obtained with a discretized sine wave. 

Note that we have found the wave shape leading to maximal flow rate, considering $N$ pumps actuated at a fixed amplitude $\varepsilon$, and with a pump length scaling as $1/N$. Numerical results indicate that, under these constraints, the energy needed to actuate $N$ pumps with the optimal waveform is larger than for a sinusoidal wave with the same amplitude of pumping $\varepsilon$, see Supp. Fig.~4a in~\cite{Note1}. For a given energy, and relaxing the constraint on the amplitude of pumping, a sinusoidal wave leads to a larger flow than our optimal wave form, see Supp. Fig.~4b in ~\cite{Note1}. We plan to investigate in the future what the optimal phases and amplitudes of actuation are in a system of $N$ pumps, under the constraint of a fixed energy.

Finally, the results presented here are valid for a system of $N$ pumps actuated independently of each other, in a regime where the imposed actuation is instantaneously transmitted to the fluid. They do not take into account effects due to the elasticity of the tubing enclosing the fluid. Elasticity introduces a compliance in the system as mentioned above, and can also lead to two neighboring pumps not being independent of each other: if the pumps are too close to each other, closing one pump will open the neighboring one. The analysis of such intricate effects is left for future work.

\begin{acknowledgments}
We thank Rom\'{e}o Antier for his contribution to the early stages of this work, and Caroline Frot for her valuable help with fabrication.
\end{acknowledgments}

\bibliography{biblio}

\end{document}